\documentclass[twocolumn]{aastex631}

\usepackage{lipsum}
\usepackage{amsmath}
\usepackage{mathtools}
\usepackage{soul}
\usepackage[nottoc]{tocbibind}

\newcommand{\be}{\begin{equation}}
\newcommand{\ee}{\end{equation}}
\shorttitle{SBO}
\shortauthors{Menon et al.}
\begin{document}
\title{UV SIGNATURES OF MAGNETAR FORMATION AND THEIR CRUCIAL ROLE FOR GW DETECTION} 
\correspondingauthor{ssmenonoffl@gmail.com}
\author{Sandhya S. Menon}
\author[0000-0002-7349-1109]{Dafne Guetta}
\affil{Physics Department,Ariel University, Ariel, Israel}
\author[0000-0003-4366-8265]{Simone Dall'Osso}
\affil{Istituto Nazionale di Fisica Nucleare - Roma 1, Piazzale Aldo Moro 2, 00185, Roma}
\begin{abstract}
The emission from shock breakouts (SBOs) represents the earliest electromagnetic (EM) signal emitted by cataclysmic events involving the formation or the merger of neutron stars (NSs). As such, SBOs carry unique information on the structure of their progenitors and on the explosion energy.~The characteristic~SBO emission is expected in the UV range, and its detection is one of the key targets of~the ULTRASAT satellite.~Among SBO sources, we
focus on a specific class involving the formation of fast spinning magnetars in the core-collapse (CC) of  massive stars.~Fast spinning magnetars are expected to 
produce a specific signature in the early UV supernova light curve, powered by the extra spin energy quickly released by the NS. Moreover, they are considered as optimal candidates for the emission of long-transient gravitational wave (GW) signals, the detection of which requires early EM triggers to boost the sensitivity of dedicated GW search pipelines.~We calculate early supernova UV light curves in the presence of a magnetar central engine, as a function of the explosion energy, ejecta mass and magnetar parameters.~We then estimate the ULTRASAT detection horizon ($z < 0.15$) as a function of the same physical parameters, and the overall expected detection rate finding that magnetar-powered SBOs  may represent up to 1/5 of the total events detected by ULTRASAT.~Moreover, at the expected sensitivity of the  LIGO/Virgo/Kagra O5 science run, one such event occurring within 5 Mpc will provide~an ideal trigger for a GW long transient search.~Future GW detectors like the Einstein Telescope will push the horizon for joint EM-GW detections to 35-40 Mpc.

\end{abstract}
\section{Introduction}
Over the last decade, major observational efforts in astronomy have been devoted to time domain surveys  (e.g. ZTF, EUCLID,
eROSITA, SKA, VRO), 
aimed~at exploiting the great discovery potential of explosive transients produced in cosmic cataclysms.~Many~among such events are expected to produce a characteristically bright emission in the UV band \citep{sagiv2014science, kulkarni2021science}, a poorly covered spectral range in the multi-wavelength monitoring of astrophysical transients.~The Ultraviolet Transient Astronomy Satellite (ULTRASAT), scheduled for launch~in 2026, is a small mission dedicated to time domain observations in the NUV band, and will carry out the~first wide-field survey of transient and variable sources, with a much larger horizon than previous UV satellites.~A key 
goal of ULTRASAT is the early~(hrs) detection of hundreds of core-collapse supernovae~(CCSN) and the high cadence (minutes) monitoring of their UV light curves.~Approximately 10\% of CCSNe  are expected to lead to the birth of magnetars \citep{beniamini2019formation}, {\it i.e.}~highly magnetic neutron stars (NS) thought to be born with millisecond spin (\citealt{duncan1992formation, thompson1993neutron}; see also \citealt{dall2021millisecond} and references therein).~Due to~these unique properties, newly formed magnetars~can be powerful machines capable
of injecting large amounts of energy and driving a shock through the expanding ejecta.~Once this shock breaks out of the ejecta a characteristic UV signature is expected in the early supernova light curve, {\it i.e.} a magnetar-driven shock breakout\footnote{This is different from, and later than, the `typical' SBO expected when the collapse-driven shock first breaks out of the progenitor star.~ULTRASAT may in principle observe both if a magnetar is formed.} (SBO; e.g.~\citealt{kasen2010supernova, leloudas2012sn,nicholl2015diversity,kasen2016magnetar, liu2021magnetar}).~The launch of ULTRASAT thus represents an ideal opportunity for the first identification of newly born magnetars, which will
settle a long-standing debate about their origin and role in various astrophysical transients. 

Moreover, newly born magnetars are expected to~emit long-lasting GW transients in the first
hours after~their birth~\citep{cutler2002gravitational, dall2009early, corsi2009gamma, dallo15, dall2018neutron,
lander2020magnetar, sur2021gravitational}.~Such GW signals can~be much stronger than that from the core-collapse itself and therefore detectable up to larger distances.~Thus,~even though magnetars occur only in $\sim$ 10\% of CCSNe, they can be detected in GWs with a larger event rate if the search horizon is $\gtrsim 2-3$ times that of CCSNe. Such an horizon can indeed be achieved and even exceeded,~at the sensitivity of the LIGO/Virgo/Kagra~(LVK) O4-O5 science runs, in the presence of an 
electromagnetic (EM) trigger\footnote{During the O2 science run, with a $\sim 3$ times lower sensitivity than in O4, GW searches for a post-merger remnant in GW 170817 reached an horizon $\lesssim 1$ Mpc \citep{abbott2019search}.}, 
e.g. the magnetar-driven SBO that ULTRASAT will provide. 


In this paper we assess the detectability of magnetar-powered SBOs with ULTRASAT.~Building 
on existing models \citep{kasen2016magnetar,liu2021magnetar}, in Sec.~\ref{sec:modelling} we calculate the SBO peak luminosity, photosferic radius and peak temperature as a function of the relevant NS parameters (spin period, $P$, and dipole magnetic field, $B$).~In Sec.~\ref{sec:horizon}~we translate these parameters into an ULTRASAT detection horizon, given the detector's sensitivity in the 0.22-0.28~$\mu$m window.~In Sec.~\ref{sec:discussion} we estimate, by exploring different parameter distributions within the extragalactic magnetar population, the implied ULTRASAT detection rate of magnetar-powered SBOs, and the prospects for joint detections of the associated GW long-transients.~We summarize our conclusions in Sec.~\ref{sec:conclusions}.
\section{Magnetar-driven SBO light curves}
\label{sec:modelling}
A newly born millisecond spinning magnetar, with angular velocity $\Omega=2\pi/P$ and moment of inertia $I$, has the rotational energy 
\begin{equation}
   E_{\rm m} = \frac{1}{2} I \Omega^2 \approx 3 \times 10^{52}~{\rm erg}~P^{-2}_{\rm ms} M_{1.4} R_{12}^2 \, ,
   \label{eq:1}
    \end{equation}
   expressing the spin period in milliseconds, the mass in units of 1.4 M$_{\odot}$ and setting the radius $R=12$ km.
The 
 NS spins down due to magnetic dipole radiation on the timescale \citep{spitkovsky2006time}
\begin{equation}
     t_m=\frac{Ic^2}{2 \mu^2 \Omega^2} 
     \approx 7 \times 10^4~{\rm s}~ \left(\frac{P_{\rm ms}}{B_{14}}\right)^2 \frac{M_{1.4}}{R^4_{12}}\, ,
    \label{eq:2}
    \end{equation}
for an aligned rotator (dipole B-field in units of $10^{14}$~G), releasing rotational energy at the rate 
\begin{equation}
    L_m =
    \frac{E_m/t_m }{(1+t/t_m)^2} \sim 10^{47}~{\rm erg~s}^{-1} \displaystyle\frac{E_{m, 52}}{ t_{m,5}}\left(1+\displaystyle t/t_m\right)^{-2} \, .
    \end{equation}
The released energy inflates a high-pressure bubble of relativistic particles and magnetic field that sweeps the  previously launched SN ejecta into a thin shell, driving a shock through~it.~Here we model the physics of this interaction according to the prescription by Kasen et al. (2016), adopting a characteristic explosion energy of $10^{51}$ erg\footnote{Small deviations from this fiducial value would not affect the model in significant ways.~In the future we plan to address the more general cases of much lower, or much larger, explosion energies, which would impact some of the model prescriptions adopted here.}.~The 
 \begin{figure*}[ht]
        \centering
        \includegraphics[scale=0.58]{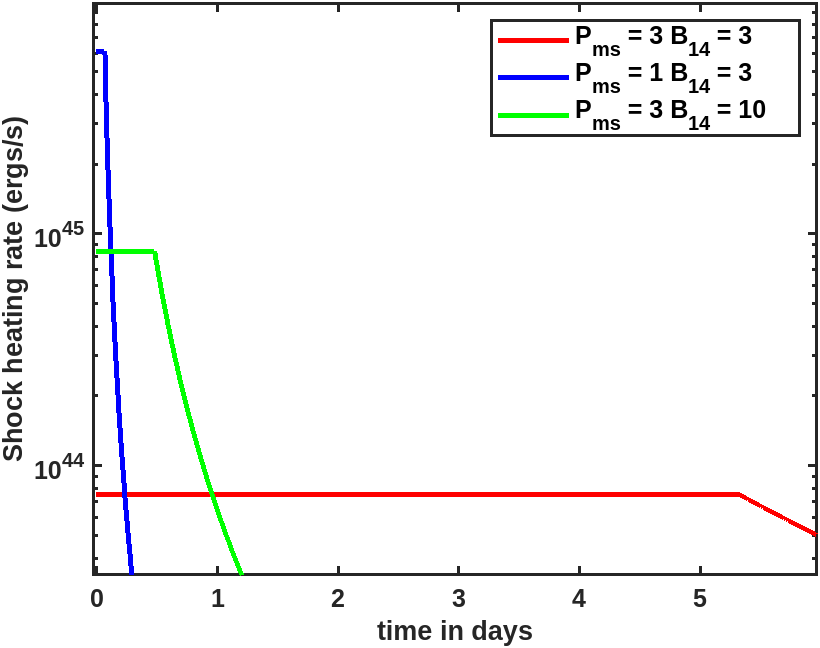}
        \includegraphics[scale=0.59]{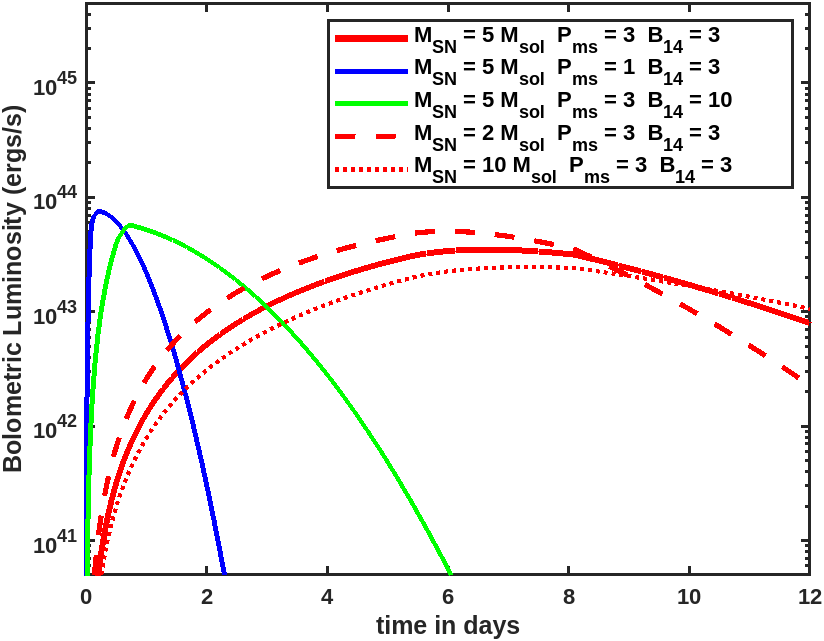}
        \caption{{\it Left Panel:} Shock heating rate, $\dot{\epsilon}_{\rm sh}$ vs. time (Eq.~\ref{eq:epsilonrate}) for fiducial parameter values of the SN ejecta ($E_{SN} = 10^{51}$~erg, $M_{SN} = 5 M_\odot$) and 
        three sets of ($P, B$)-values for the magnetar central engine, as indicated in the legend.~{\it Right Panel:} Bolometric SBO light curve (Eq.~\ref{eq:LSBO-bolo}) for the same set of parameters as in the left panel.~The dashed and dot-dashed red curves are for the same magnetar as the solid red curve, but for different ejecta masses:  2 $M_\odot$ (dashed) or 10 $M_\odot$ (dot-dashed).~The impact of changing the ejecta mass is substantially smaller than that of different magnetar parameters, to which our model is much more sensitive.~In particular, while the (bolometric) peak luminosity has a moderate dependence on the magnetar spin and $B$-field, the light curve evolution and peak time are strongly dependent on these parameters. For reference the peak time of the blue, green and red solid curves are, respectively, $\approx 0.2, 0.7$ and 5.5 days.}
        \label{fig:my_label1}
\end{figure*}
density profile within the ejecta shell can be approximated by a broken power-law, which is shallow in the inner region and becomes very steep close to the surface,
\begin{equation}
 \rho(r,t)=
\begin{cases}
\zeta_\rho \displaystyle \frac{M_{SN}}{v_t^3 t^3} \left(\displaystyle \frac{r}{v_t t}\right)^{-\delta}   & ~~ \text{if $v<v_t$} \\
\zeta_\rho \displaystyle \frac{M_{SN}}{v_t^3 t^3} \left(\displaystyle\frac{r}{v_t t}\right)^{-n}  & ~~\text{if $v\geq v_t$} \, , \\
\end{cases} 
\end{equation}
with the transition occurring at the velocity coordinate
\begin{equation}
   v_t 
   \approx 3 \times 10^8~{\rm cm~s}^{-1} \zeta_v \left(\displaystyle \frac{E_{SN, 51}}{M_{SN}/5 M_\odot}\right)^{1/2}
     \label{eq:6}
    \end{equation} 
    for an homologous expansion of the shell.~Typical values are $\delta =1$ and $n=10$ for CCSNe \citep{chevalier1989asymmetric}, while 
   \begin{equation}
    \zeta_v= \frac {2 (5-\delta)(n-5)}{(n-3)(3-\delta)} ^{1/2}
    \label{eq:5}
    \end{equation}
    \begin{equation}
    \zeta_\rho= \frac {(n-3)(3-\delta)}{4\pi(n-\delta)} \, .
    \label{eq:13}
    \end{equation}
By using the mass, momentum, and energy equations for the shock \citep{kasen2016magnetar}, we can determine its radius $r_{\rm s}$ as a function of time
 \begin{equation}
     r_{\rm s}(t)= v_t t_{\rm tr}^{1-\alpha}t^\alpha
     \label{eq:4}
    \end{equation}
    \begin{equation}
 \alpha=
\begin{cases}
5/4 & \text{if $v<v_t$} \\
3/2 & \text{if $v\geq v_t$} \, ,
\end{cases}
\end{equation}
where $t_{\rm tr}$ is the time it takes for the shock to propagate through the inner ejecta, 
 \begin{equation}
     t_{\rm tr}= \zeta_{\rm tr} (E_{SN}/E_m) t_m
     \label{eq:7}
    \end{equation}
 \begin{equation}
    \zeta_{\rm tr}= \frac {2 (n-5)(9-2\delta)(11-2\delta)}{(5-\delta)^2(n-\delta)(3-\delta)}  \, .
    \label{eq:9}
    \end{equation}
Inside the shock, energy is dissipated at the rate  $\dot{\epsilon}_{\rm sh}$ 
   \begin{equation}
   \label{eq:epsilonrate}
   \dot{\epsilon}_{\rm sh}= 4 \pi r_{\rm s}^2 \frac{\rho}{2} v_{\rm ej}^3 \eta^3 \, ,
    \end{equation} 
where $v_{\rm ej} = r_{\rm s} /t$ is the ejecta velocity at radius $r_{\rm s}$ and $\eta$ is the shock strength parameter
  \begin{equation}
      \eta(t) = \frac{v_{\rm s}-v_{\rm ej}}{v_{\rm ej}} = \frac{t}{r_{\rm s}} \left(\frac{dr_{\rm s}}{dt}-\frac{r_{\rm s}}{t}\right) \, ,
      \label{eq:11}
  \end{equation}

$v_s$ being the shock velocity.~The shock will become radiative once it reaches the region within the ejecta where the photon optical depth $\tau \approx c/v_s$. This shock breakout (SBO) happens at radius 
\begin{equation}
    r_{\rm bo} = v_t t \left(\displaystyle \frac{t_d }{\sqrt{n-1}   t}\right)^{2/(n-2)} \, ,
    \label{eq:rbo}
\end{equation}
where $t_d$ is the effective diffusion time, 
\begin{equation}
    t_d = \sqrt{\frac{\zeta_\rho \kappa M_{SN}} {v_t c}} \, ,
\end{equation}
and $\kappa = 0.1~{\rm cm}^2~{\rm g}^{-1}$  the opacity of the ejecta material. \\
Equating the breakout radius $r_{\rm bo}$ to the shock radius $r_s$, we obtain the SBO time, 
\begin{equation}
    t_{\rm bo}\approx \left[\frac{\zeta_{\rm tr}^{1-\beta}}{(n-1)^{\beta/2}}\right] t_d^\beta t_m^{1-\beta} (E_{SN}/E_m)^{1-\beta}
    \label{eq:tbo}
\end{equation}
with
\begin{equation}
    \beta= \frac {2}{(\alpha -1)(n-2)+2}  \, .
    \label{eq:15}
    \end{equation}
Most of the radiation will escape from the region where the photon optical depth $\tau \approx 1$, which defines the photospheric radius,  
 \begin{equation}
 \label{eq:photospheric}
 r_p = 1.2 v_t t_d \left(\frac{t_{\rm bo}}{t_d}\right)^{(n-3)/(n-1)} \, .
 \end{equation}

The resulting luminosity evolution is \citep{kasen2016magnetar}
\begin{equation}
     \label{eq:LSBO-bolo}
     L_{\rm sbo} (t) \approx e^{-\left(\displaystyle \frac{t}{t_{\rm bo}}\right)^2}   \int_0^{t} 2\dot{\epsilon}_{sh}(t') \frac{t' }{t_{\rm bo}^2} e^{\left(\displaystyle \frac{t'}{t_{\rm bo}}\right)^2} \,dt'\, . 
    \end{equation} 
Fig.~\ref{fig:my_label1} depicts the time evolution of the shock heating rate (left panel) and of the SBO bolometric luminosity (right panel) for selected values of the magnetar spin period and magnetic dipole field (see caption for details). We 
    \begin{figure*}
        \centering
        \includegraphics[scale=0.415]{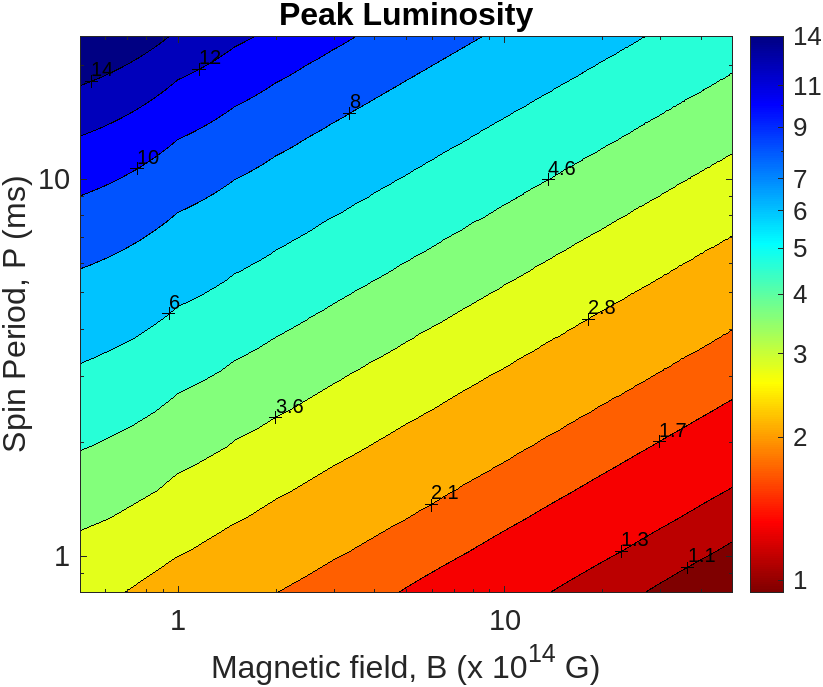}
        \includegraphics[scale=0.415]{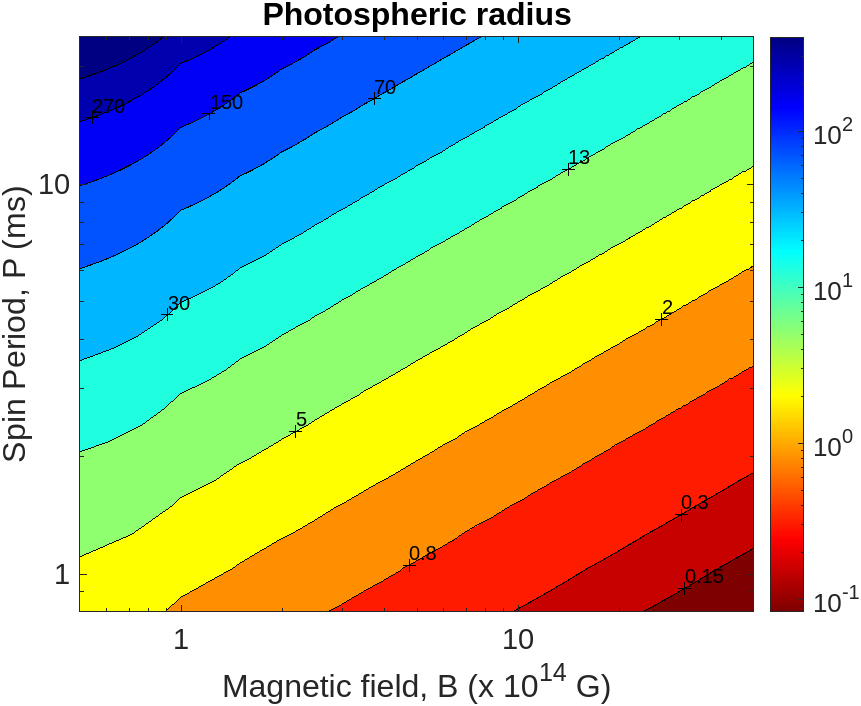}
        \includegraphics[scale=0.415]{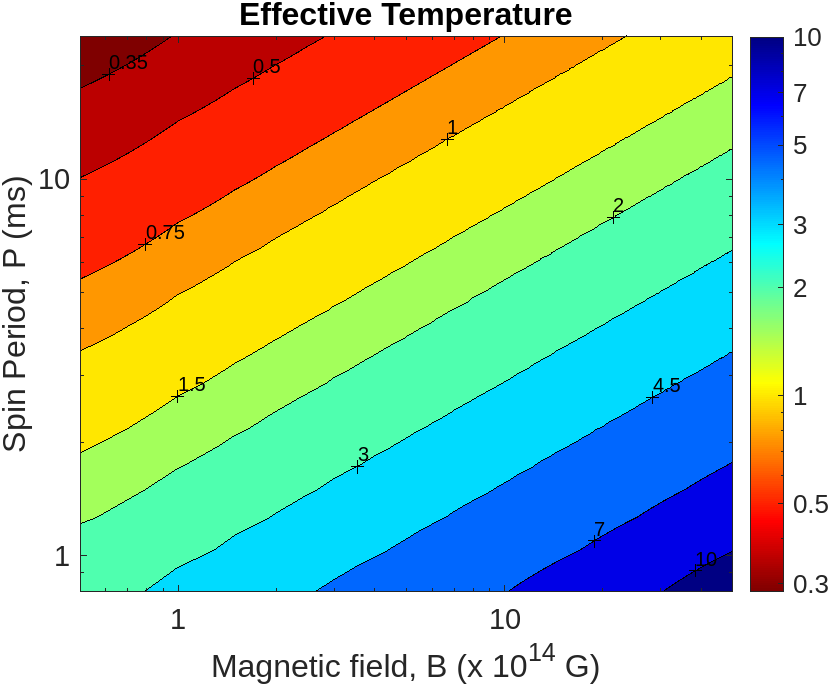}
        \caption{{\it Left Panel:} the bolometric SBO peak luminosity, $L_{\rm SBO}^{\rm peak}$ in units of $10^{43}$ erg s$^{-1}$, as a function of the magnetar parameters ($B, P$; eq.~\ref{eq:cL-SBO-peak}), for fixed ejecta parameters, $E_{SN} =10^{51}$ erg and $M_{SN} = 5 M_\odot$; {\it Middle Panel:} 
         Photospheric radius, $r_p$ in units of $10^{14}$~cm, as a function of the magnetar parameters $(B, P)$, and the same ejecta properties as in the left panel; {\it Right Panel:} Effective blackbody temperature, $T_{\rm bb}$ in units of $10^4$ K, vs. magnetar parameters as derived from the relation $L_{\rm SBO}^{\rm peak} = 4 \pi \sigma r^2_p T^4_{\rm bb}$, for the same ejecta properties as in the previous panels.}
            \label{fig:treplot}
\end{figure*}
verified that the SBO lightcurve peak corresponds approximately to the shock heating rate at
the SBO time, $t_{\rm bo}$, and can therefore be expressed as
 \begin{equation}
    L_{\rm sbo}^{\rm peak} \approx 9 \frac{E_{SN}}{t_d} \eta^3 \left( \frac{t_m}{t_d} \frac{E_{SN}}{E_m}\right)^{(n-8) (1-\beta)/(n-2)} \, .
    \label{eq:cL-SBO-peak}
    \end{equation}
\section{PEAK LUMINOSITY IN THE ULTRASAT BAND AND DETECTION HORIZON}
\label{sec:horizon}
The amount of SBO radiation going into the ULTRASAT observing band ($\lambda_{\rm min} = 220$ nm; $\lambda_{\rm max}~=280$~nm) can be estimated from the expected blackbody emission spectrum.~Formally, the  
SBO peak luminosity in the observed UV band, $L_{UV}^{\rm peak}$, can be expressed as a fraction $f_{UV}$ of $L_{\rm sbo}^{\rm peak}$
where $f_{UV}$
 \begin{equation}
 \label{eq:fuv}
    f_{UV}= \frac{\displaystyle \int_{\lambda_{\rm min}/(1+z)}^{\lambda_{\rm max}/(1+z)} B_\lambda(T_{\rm bb}) \,d\lambda\ }{{\sigma T_{\rm bb}^4}} \, ,
    \end{equation} 
$B_\lambda$ is the spectral radiance per unit wavelength, $\sigma$~the Stefan–Boltzmann constant and $T_{\rm bb}$ the blackbody temperature, which is determined by the bolometric peak luminosity and the photospheric radius.~The latter quantities depend, according to eqs.~\ref{eq:cL-SBO-peak} and \ref{eq:photospheric}, on the magnetar parameters ($B, P$) as well as on the SN parameters ($E_{SN}, M_{SN}$).~Note that, in eq.~\ref{eq:fuv}, the observed spectral range must be calculated at the source, {\it i.e.} taking into account the cosmological  redshift.~Thus, $f_{UV}$ is a function of $z$ as well as of the source physics parameters.

Correspondingly, we can derive a maximum redshift, $z_{\rm max}$, for detection as a function of the blackbody temperature, given the ULTRASAT mean limiting magnitude of 22.4 ABmag in the 220-280 nm band \citep{ultrasat}.~This maximum redshift can then be expressed as a function of $(B, P, E_{\rm SN}, M_{\rm SN})$ where, as illustrated in the right panel of Fig.~\ref{fig:my_label1}, the dependence of $L_{\rm sbo}^{\rm peak}$ - hence of $z_{\rm max}$ - on the ejecta mass is mild at most. 

Up to this point, our estimate does not account for~the effects of extinction in the host galaxy\footnote{The minor extinction within our Galaxy is discussed, e.g., in \citep{ganot2016detection}.}.~While these are expected to present wide variations among different events, we adopt an average value of $A_{NUV} \sim 1.75$~mag based on past UV observations of similar sources with $Swift$/UVOT or GALEX \citep{ganot2016detection}.~  

Fig.~\ref{fig:treplot} shows our model results as contour plots of $L_{\rm sbo}^{\rm peak}$, $r_p$ and $T_{\rm bb}$~vs.~($B, P$), for fixed parameters of the SN ejecta, $E_{\rm SN} = 10^{51}$ erg and $M_{\rm SN} =5 M_\odot$.~Fig.~\ref{fig:zmax} depicts the value of $z_{\rm max}$ as a function of the magnetar parameters, for the same SN ejecta properties, with (left panel) or without (right panel) extinction.  
\section{EXPECTED EVENT RATE AND MULTI-MESSENGER IMPLICATIONS}
\label{sec:discussion}
\begin{figure*}
        \centering
        \includegraphics[scale=0.4]{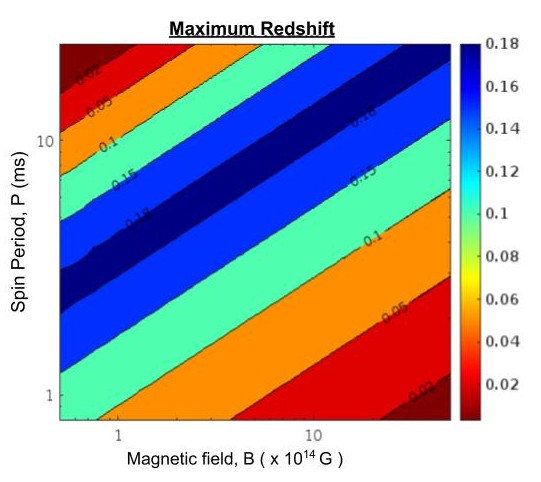}
        \includegraphics[scale=0.41]{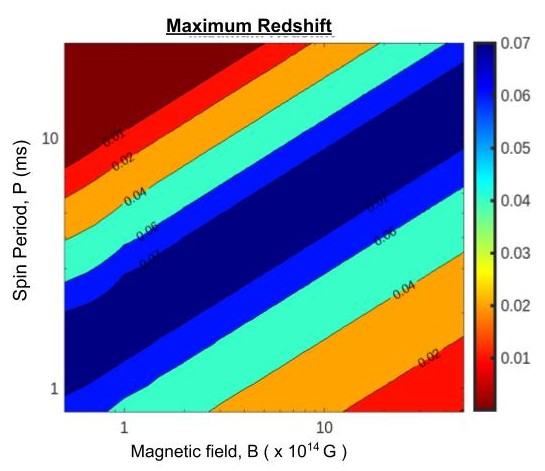}
        \caption{The maximum redshift,  $z_{\rm max}$, for detection~with ULTRASAT as a function of the magnetar parameters ($B, P$), for fiducial ejecta properties $E_{SN} =10^{51}$ erg, $M_{SN} = 5 M_\odot$.~{\it Left Panel:} theoretical predictions without UV extinction; {\it Right Panel:} as in the left panel, but including an average extinction $A_{NUV} = 1.75$~mag.}
            \label{fig:zmax}
\end{figure*}

Our results indicate that magnetar-driven SBOs can be detected by ULTRASAT out to a distance $D_{\rm L, max} \approx 300$ Mpc for an extinction $A_{NUV} =1.75$~mag, with a significant dependence on the B-field and spin energy of the NS. In order to estimate an expected rate of detections we must therefore average the $z_{\rm max}$ distribution (Fig.~\ref{fig:zmax})~over the probability distribution of magnetar parameters within the population.~To this goal, we will use both the un-extinguished and the extinguished values of $z_{\rm max}$ vs. $(B, P)$, in order to provide a measure of the impact of extinction.

The $B$- and $P$- probability distributions of magnetars at birth are currently unknown.~Thus, we will adopt agnostic distributions for both parameters,~{\it i.e.} a constant in logarithmic intervals,  
   $ {\rm I\kern-0.15em P}(B) =
   k_B/B$ 
   and  $ {\rm I\kern-0.15em P}(P) = \displaystyle k_P/P$.~Later we will check the dependence of our results on this assumption, by changing this prior.~With  
these provisions, the expected number of events per year that can be detected with ULTRASAT is expressed as 

\begin{equation}
\begin{split}
    \dot{N}_{\rm ev} = \displaystyle \int\displaylimits_{P_{\rm min}}^{P_{\rm max}}d P \int\displaylimits_{B_{\rm min}}^{B_{\rm max}} dB \int\displaylimits_{0}^{z_{\rm max}(B,P)} \frac{{\cal R}(z)}{(1+z)} \frac{dV}{dz}
   {\rm  I\kern-0.15em P}(B) {\rm I\kern-0.15em P}(P) dz
    \end{split}
    \label{eq:R_mean}
    \end{equation}
where ${\cal R}(z)$ is the comoving volumetric
rate of SNe~that lead to the formation of a magnetar, which we take to~be $\sim 0.1$ of the rate of CCSNe \citep{gaensler2005, beniamini2019formation}.~The $(1+z)$ factor in the denominator accounts for the time dilation of the rate.~The latter is given by \citep{madau2014cosmic}
\begin{equation}
    {\cal R}_{CCSN} (z) = k_{\rm CC} \times \psi(z)
    \end{equation} 
    \begin{equation*}
    \psi(z) = 0.015 \frac{(1 + z)^{2.7}}{1 + \left( \displaystyle \frac{1 + z}{2.9} \right) ^{5.6}} \rm \,~M_\odot \, year^{-1} Mpc^{-3} \, ,
    \end{equation*}
    where $k_{\rm CC} = 0.0068 \rm M_\odot^{-1}$ is the number of stars that explode as SNe per unit mass, for a Salpeter IMF with $\rm M_{min} = 8 M_\odot$ and $\rm M_{max} = 40 M_\odot$ enclosing the stellar mass range that will lead to the formation of a NS. \\
The co-moving volume element $dV(z)$ is 
\begin{equation}
    dV(z) = D^3_H \frac{d^2_C (z)}{F(z)}\, dz\, d\Omega  \, ,
    \label{dV}
    \end{equation}  
 where $D_H = c/H_0$ is Hubble distance, $d_C(z)$ the proper distance at redshift $z$, 
\begin{equation}  d_C(z) =  \int\displaylimits_0^{z} \frac{1}{F(z')} \,dz' ,
 \end{equation}
and  \begin{equation*}
    F(z)= \sqrt{\Omega_M(1+z)^3+\Omega_k(1+z)^2+\Omega_{\Lambda}} \, ,
 \end{equation*}
 with the cosmological parameters $\Omega_M =0.685, \Omega_\Lambda=0.315$, $\Omega_k=0$ \citep{aghanim2020planck}. 
     
Plugging everything into eq.~\ref{eq:R_mean} we finally obtain
\begin{equation}
\begin{split}
  \dot{N}_{\rm ev}= 4 \pi k_{CC}k_B k_P \left(\frac{c}{H_0} \right) ^3 \displaystyle \int\displaylimits_{P_{\rm min}}^{P_{\rm max}}\frac{dP}{P} \int\displaylimits_{B_{\rm min}}^{B_{\rm max}}\frac{dB}{B} \times \\ \int\displaylimits_{0}^{z_{\rm max}(B,P)} 
      \displaystyle \frac{\psi(z) d^2_C(z)}{(1+z) F(z)}\, dz    \, ,
    \end{split}
    \label{eq:expectedrate}
    \end{equation} 
where $k_B = \left[\log \left(B_{\rm max}/B_{\rm min}\right)\right]^{-1} \approx 0.217$ and $k_P =  \left[\log \left(P_{\rm max}/P_{\rm min}\right)\right]^{-1} \approx 0.294$. 

Choosing the range of relevant spin periods 0.8-24~ms and of dipole B-fields $(0.5-50)\times 10^{14}$ G, and limiting our calculation to the ULTRASAT high-cadence ($\sim 300$ s) survey with a field-of-view $\sim 204$ deg$^2$ \citep{ultrasat23},~eq.~\ref{eq:expectedrate} gives $\dot{N}_{\rm ev} \approx (3-30)$~yr$^{-1}$,  depending on the amount of extinction ($A_{NUV} = 0-1.75$~mag) and on the ejecta mass ($M_{SN} = 5-10 M_\odot$).~While we have assumed that $\sim$~10\% of CCSNe can form a magnetar,~their detectability is increased over `typical' SBO sources if~the average $A_{NUV} \lesssim 0.8$~mag, leading to a larger expected fraction ($\sim 10-20$\%) of magnetar-powered SBOs among ULTRASAT detections.~If, on the other hand, $A_{NUV} = 0.8-1.75$~mag, we expect $\sim 2-10$~\% of such detections to include a magnetar central engine.

While millisecond spinning magnetars have been proposed as central engines in Super Luminous Supernovae (SLSNe) by several authors (e.g. \citealt{kasen2010supernova, Woosely2010, mazzali14, nicholl2015, kasen2016magnetar}), the model discussed here does not necessarily imply that a SLSN should be associated to a magnetar-driven SBO.~Actually, the rate of SLSNe is estimated to be orders of magnitude lower than the  magnetar birth rate (\citealt{Quimby2013}), and 
indeed represents an issue for {\it any} progenitor model, likely pointing to the existence of additional conditions - besides a powerful central engine - to be met in order for a SLSN to occur.~Some of these conditions were already discussed (e.g. \citealt{metzger15, kasen2016magnetar}); we add here, and discuss further below, that a dominant GW spindown in the early hours after a magnetar birth may provide an additional, straightforward interpretation for this rate mismatch (see, e.g., \citealt{dallosso2007, dall2009early, dallo15, dall2018neutron, dall2021millisecond}).

To evaluate the robustness of our prediction, we repeated the estimate by adopting a gaussian prior for~the $B$-distribution, peaking at $B_{0,14}=5$ and with a $\sigma_{B, 14}=2.5$ in order to reproduce the $B$-range for known magnetars within a 3$\sigma$ interval.~Because we expect a spin period distribution which is peaked around a few milliseconds, we maintain the log-uniform distribution as~the best prior representation for $P$.~The expected detection rate for magnetar-powered SBOs becomes in this case $\sim (2-20)$ yr$^{-1}$, showing that our prediction is robust with respect to the prior. 

The NS spin and B-field play a key role in shaping the light curves of magnetar-driven SBOs, as shown in sec.~\ref{sec:modelling}.~ULTRASAT detections  
can therefore provide crucial evidence for the presence of a magnetar central engine, and constrain its spin energy and magnetic field.~Other model parameters that may affect the light curve shape, peak luminosity and overall detectability are the explosion energy and ejecta mass.~In this first study we have assumed fiducial values for both parameters, and showed that  the ejecta mass may only have a moderate impact on the lightcurve peak (Fig. 1, right panel), hence on the detectability of the UV SBO.~In a future study we will also quantify in greater detail the implications of a wide range of explosion energies, which may be more subtle when substantial deviations from fiducial values are considered.

Moreover, magnetar-driven SBOs can provide crucial EM triggers for GW searches of long-transients emitted by newly formed magnetars.~They will help enhance the sensitivity of GW searches in two complementary ways: on the one hand by providing the start time for the transient signal, and on the other hand by constraining the NS spin period and B-field, which in turn determine the GW signal shape.   

A previous search for this type of GW signals was carried out during the O2 science run of LIGO/Virgo, looking for the merger remnant in GW 170817 \citep{abbott2019search}.~While this led to a non-detection, it showed that the existing pipelines were capable of reaching an horizon $\lesssim$ 1~Mpc.\\ At the improved sensitivity expected in O5, this projects to at least 4 Mpc, likely be extended farther by constraints on the magnetar parameters provided by the SBO light curve (and by foreseen upgrades to pipeline performances).~The exact impact of these factors on the detection horizon will be quantified in a dedicated study. 

The possibility to extend the search horizon for GW long-transients to beyond 4 Mpc leads to the expectation of at least $0.1$ events per yr (e.g. \citealt{kistler2013tomography}) during the O5 science run.~In the future, the greatly improved sensitivity of third generation detectors like the Einstein Telescope will extend the search horizon by a factor $\gtrsim 7$,  and the corresponding event rate by $> 10$ \citep{dall2018neutron, dall2021millisecond}.
\section{CONCLUSIONS AND OUTLOOK}
\label{sec:conclusions}
In this work we have adopted a model for magnetar-driven SBOs that has been long discussed in the literature \citep{kasen2016magnetar, liu2021magnetar}.~We have illustrated the main physical properties of the model, calculating the SBO lightcurve, the peak luminosity, photospheric radius and effective blackbody temperature of the expected emission as a function of the magnetar magnetic field and spin period. This enables two goals, namely (a)~predicting~the ULTRASAT detection rate of such events, which we find to be $\sim 2-30$ yr$^{-1}$,  mostly determined by the~amount~of extinction and regardless of the (unknown) $B-$~and $P$-distributions within the population and (b)~fitting~observations with model light curves that will constrain the magnetar spin and magnetic field, as well as the explosion energy and ejecta mass.~Note that, even in the case of non-detections, the model will allow to put stringent constraints on scenarios of magnetar formation.

The calculations presented in this work are amenable to further developments, particularly when considering the possiblity of different stellar progenitors (e.g. RSGs, BSGs, WR stars) leading to different explosion energies and/or ejecta masses \citep{ganot2016detection}.~Additionally, the thermalization efficiency of the magnetar wind inside the shock~will be taken into account as it can affect the SBO detectability in the NUV band \citep{kasen2016magnetar}.~An examination of these effects in the context of a detailed observing strategy will be explored in a forthcoming paper. 

Moreover, we have discussed the great relevance that magnetar-driven SBOs detected by ULTRASAT will have for the joint search of GW long transients emitted by newly born magnetars.~Not only will magnetar-driven SBOs represent a necessary EM trigger for GW searches, but they will also provide direct constraints on the GW signal parameters.~Both factors will contribute to improving the sensitivity of GW searches, hence extending their horizon and the expected event rate. 


It is interesting to further note that the Zwicky Transient Facility (ZTF) telescope will be able to observe most of the SNe associated to this kind of events.~For instance,   
\cite{Guetta2020} find that CCSNe can be detected out to a maximum distance of $D_{\rm max} \sim 200$ Mpc by the ZTF, implying hundreds of CCSN detections per year.~Since the ZTF horizon is comparable to the estimated ULTRASAT horizon for magnetar-driven SBOs ($\approx 300$ Mpc for a fiducial extinction of $\sim 2$ mag), we expect most of the ULTRASAT events to have a counterpart in the ZTF, thus enabling prompt ground-based follow-ups of ULTRASAT sources.~Conversely, the ZTF will observe hundreds of CCSNe that are out of the high-cadence field of view of ULTRASAT.~The latter events may thus trigger ToO observations with ULTRASAT, further enhancing its detection prospects.~

Within the estimated GW horizon ($\gtrsim 4$ Mpc) for the O5 science run, we expect ULTRASAT to easily detect all magnetar-driven SBOs, implying a multi-messenger detection rate $\gtrsim 0.1$ yr$^{-1}$.~A 7-8 times larger horizon could be reached by the Einstein Telescope, which would include the whole Virgo Cluster and possibly beyond it.~Again, ULTRASAT would be able to easily detect any SBOs within this distance.~Therefore, with a minimum estimated magnetar birth rate of $\sim$ 1 yr$^{-1}$ within 30~Mpc (e.g. \citealt{stella2005gravitational, dall2009early}), we expect a minimum rate of multi-messenger detections of 1 magnetar per yr. 

\section*{acknowledgements}
SD acknowledges funding from the European Union’s Horizon2020 research and innovation programme under the Marie Skłodowska-Curie (grant agreement No. 754496). SD
thanks the University of Ariel for their hospitality during the visit and creative atmosphere
We thank Prof. Pia Astone for her suggestions.


\bibliographystyle{aasjournal}
\bibliography{Reference}

\begin{thebibliography}{}
\expandafter\ifx\csname natexlab\endcsname\relax\def\natexlab#1{#1}\fi
\providecommand{\url}[1]{\href{#1}{#1}}
\providecommand{\dodoi}[1]{doi:~\href{http://doi.org/#1}{\nolinkurl{#1}}}
\providecommand{\doeprint}[1]{\href{http://ascl.net/#1}{\nolinkurl{http://ascl.net/#1}}}
\providecommand{\doarXiv}[1]{\href{https://arxiv.org/abs/#1}{\nolinkurl{https://arxiv.org/abs/#1}}}

\bibitem[{Abbott {et~al.}(2019)Abbott, Abbott, Abbott, Acernese, Ackley, Adams,
  Adams, Addesso, Adhikari, Adya, {et~al.}}]{abbott2019search}
Abbott, B.~P., Abbott, R., Abbott, T., {et~al.} 2019, The Astrophysical
  Journal, 875, 160

\bibitem[{Aghanim {et~al.}(2020)Aghanim, Akrami, Ashdown, Aumont, Baccigalupi,
  Ballardini, Banday, Barreiro, Bartolo, Basak, {et~al.}}]{aghanim2020planck}
Aghanim, N., Akrami, Y., Ashdown, M., {et~al.} 2020, Astronomy \& Astrophysics,
  641, A6

\bibitem[{Beniamini {et~al.}(2019)Beniamini, Hotokezaka, van~der Horst, \&
  Kouveliotou}]{beniamini2019formation}
Beniamini, P., Hotokezaka, K., van~der Horst, A., \& Kouveliotou, C. 2019,
  Monthly Notices of the Royal Astronomical Society, 487, 1426

\bibitem[{Chevalier \& Soker(1989)}]{chevalier1989asymmetric}
Chevalier, R.~A., \& Soker, N. 1989, Astrophysical Journal, Part 1 (ISSN
  0004-637X), vol. 341, June 15, 1989, p. 867-882., 341, 867

\bibitem[{Corsi \& M{\'e}sz{\'a}ros(2009)}]{corsi2009gamma}
Corsi, A., \& M{\'e}sz{\'a}ros, P. 2009, The Astrophysical Journal, 702, 1171

\bibitem[{Cutler(2002)}]{cutler2002gravitational}
Cutler, C. 2002, Physical Review D, 66, 084025

\bibitem[{{Dall'Osso} {et~al.}(2015){Dall'Osso}, {Giacomazzo}, {Perna}, \&
  {Stella}}]{dallo15}
{Dall'Osso}, S., {Giacomazzo}, B., {Perna}, R., \& {Stella}, L. 2015, \apj,
  798, 25, \dodoi{10.1088/0004-637X/798/1/25}

\bibitem[{Dall'Osso {et~al.}(2009)Dall'Osso, Shore, \& Stella}]{dall2009early}
Dall'Osso, S., Shore, S.~N., \& Stella, L. 2009, Monthly Notices of the Royal
  Astronomical Society, 398, 1869

\bibitem[{{Dall'Osso} \& {Stella}(2007)}]{dallosso2007}
{Dall'Osso}, S., \& {Stella}, L. 2007, \apss, 308, 119,
  \dodoi{10.1007/s10509-007-9323-0}

\bibitem[{{Dall'Osso} \& {Stella}(2022)}]{dall2021millisecond}
{Dall'Osso}, S., \& {Stella}, L. 2022, in Astrophysics and Space Science
  Library, Vol. 465, Astrophysics and Space Science Library, ed.
  S.~{Bhattacharyya}, A.~{Papitto}, \& D.~{Bhattacharya}, 245--280,
  \dodoi{10.1007/978-3-030-85198-9_8}

\bibitem[{Dall’Osso {et~al.}(2018)Dall’Osso, Stella, \&
  Palomba}]{dall2018neutron}
Dall’Osso, S., Stella, L., \& Palomba, C. 2018, Monthly Notices of the Royal
  Astronomical Society, 480, 1353

\bibitem[{Duncan \& Thompson(1992)}]{duncan1992formation}
Duncan, R.~C., \& Thompson, C. 1992, The Astrophysical Journal, 392, L9

\bibitem[{{Gaensler} {et~al.}(2005){Gaensler}, {McClure-Griffiths}, {Oey},
  {Haverkorn}, {Dickey}, \& {Green}}]{gaensler2005}
{Gaensler}, B.~M., {McClure-Griffiths}, N.~M., {Oey}, M.~S., {et~al.} 2005,
  \apjl, 620, L95, \dodoi{10.1086/428725}

\bibitem[{Ganot {et~al.}(2016)Ganot, Gal-Yam, Ofek, Sagiv, Waxman, Lapid,
  Kulkarni, Ben-Ami, Kasliwal, Chelouche, {et~al.}}]{ganot2016detection}
Ganot, N., Gal-Yam, A., Ofek, E.~O., {et~al.} 2016, The Astrophysical Journal,
  820, 57

\bibitem[{{Guetta} {et~al.}(2020){Guetta}, {Rahin}, {Bartos}, \& {Della
  Valle}}]{Guetta2020}
{Guetta}, D., {Rahin}, R., {Bartos}, I., \& {Della Valle}, M. 2020, \mnras,
  492, 843, \dodoi{10.1093/mnras/stz3245}

\bibitem[{Kasen \& Bildsten(2010)}]{kasen2010supernova}
Kasen, D., \& Bildsten, L. 2010, The Astrophysical Journal, 717, 245

\bibitem[{Kasen {et~al.}(2016)Kasen, Metzger, \& Bildsten}]{kasen2016magnetar}
Kasen, D., Metzger, B.~D., \& Bildsten, L. 2016, The Astrophysical Journal,
  821, 36

\bibitem[{Kistler {et~al.}(2013)Kistler, Haxton, \&
  Y{\"u}ksel}]{kistler2013tomography}
Kistler, M.~D., Haxton, W.~C., \& Y{\"u}ksel, H. 2013, The Astrophysical
  Journal, 778, 81

\bibitem[{Kulkarni {et~al.}(2021)Kulkarni, Harrison, Grefenstette, Earnshaw,
  Andreoni, Berg, Bloom, Cenko, Chornock, Christiansen,
  {et~al.}}]{kulkarni2021science}
Kulkarni, S., Harrison, F.~A., Grefenstette, B.~W., {et~al.} 2021, arXiv
  preprint arXiv:2111.15608

\bibitem[{Lander \& Jones(2020)}]{lander2020magnetar}
Lander, S., \& Jones, D. 2020, Monthly Notices of the Royal Astronomical
  Society, 494, 4838

\bibitem[{Leloudas {et~al.}(2012)Leloudas, Chatzopoulos, Dilday, Gorosabel,
  Vinko, Gallazzi, Wheeler, Bassett, Fischer, Frieman,
  {et~al.}}]{leloudas2012sn}
Leloudas, G., Chatzopoulos, E., Dilday, B., {et~al.} 2012, Astronomy \&
  Astrophysics, 541, A129

\bibitem[{Liu {et~al.}(2021)Liu, Gao, Wang, \& Yang}]{liu2021magnetar}
Liu, L.-D., Gao, H., Wang, X.-F., \& Yang, S. 2021, The Astrophysical Journal,
  911, 142

\bibitem[{Madau \& Dickinson(2014)}]{madau2014cosmic}
Madau, P., \& Dickinson, M. 2014, Annual Review of Astronomy and Astrophysics,
  52, 415

\bibitem[{{Mazzali} {et~al.}(2014){Mazzali}, {McFadyen}, {Woosley}, {Pian}, \&
  {Tanaka}}]{mazzali14}
{Mazzali}, P.~A., {McFadyen}, A.~I., {Woosley}, S.~E., {Pian}, E., \& {Tanaka},
  M. 2014, \mnras, 443, 67, \dodoi{10.1093/mnras/stu1124}

\bibitem[{{Metzger} {et~al.}(2015){Metzger}, {Margalit}, {Kasen}, \&
  {Quataert}}]{metzger15}
{Metzger}, B.~D., {Margalit}, B., {Kasen}, D., \& {Quataert}, E. 2015, \mnras,
  454, 3311, \dodoi{10.1093/mnras/stv2224}

\bibitem[{Nicholl {et~al.}(2015)Nicholl, Smartt, Jerkstrand, Inserra, Sim,
  Chen, Benetti, Fraser, Gal-Yam, Kankare, {et~al.}}]{nicholl2015diversity}
Nicholl, M., Smartt, S.~J., Jerkstrand, A., {et~al.} 2015, Monthly Notices of
  the Royal Astronomical Society, 452, 3869

\bibitem[{{Nicholl} {et~al.}(2015){Nicholl}, {Smartt}, {Jerkstrand}, {Inserra},
  {Sim}, {Chen}, {Benetti}, {Fraser}, {Gal-Yam}, {Kankare}, {Maguire}, {Smith},
  {Sullivan}, {Valenti}, {Young}, {Baltay}, {Bauer}, {Baumont}, {Bersier},
  {Botticella}, {Childress}, {Dennefeld}, {Della Valle}, {Elias-Rosa},
  {Feindt}, {Galbany}, {Hadjiyska}, {Le Guillou}, {Leloudas}, {Mazzali},
  {McKinnon}, {Polshaw}, {Rabinowitz}, {Rostami}, {Scalzo}, {Schmidt},
  {Schulze}, {Sollerman}, {Taddia}, \& {Yuan}}]{nicholl2015}
{Nicholl}, M., {Smartt}, S.~J., {Jerkstrand}, A., {et~al.} 2015, \mnras, 452,
  3869, \dodoi{10.1093/mnras/stv1522}

\bibitem[{{Quimby} {et~al.}(2013){Quimby}, {Yuan}, {Akerlof}, \&
  {Wheeler}}]{Quimby2013}
{Quimby}, R.~M., {Yuan}, F., {Akerlof}, C., \& {Wheeler}, J.~C. 2013, \mnras,
  431, 912, \dodoi{10.1093/mnras/stt213}

\bibitem[{Sagiv {et~al.}(2014)Sagiv, Gal-Yam, Ofek, Waxman, Aharonson,
  Kulkarni, Nakar, Maoz, Trakhtenbrot, Phinney, {et~al.}}]{sagiv2014science}
Sagiv, I., Gal-Yam, A., Ofek, E., {et~al.} 2014, The Astronomical Journal, 147,
  79

\bibitem[{{Shvartzvald} {et~al.}(2023{\natexlab{a}}){Shvartzvald}, {Waxman},
  {Gal-Yam}, {Ofek}, {Ben-Ami}, {Berge}, {Kowalski}, {B{\"u}hler}, {Worm},
  {Rhoads}, {Arcavi}, {Maoz}, {Polishook}, {Stone}, {Trakhtenbrot},
  {Ackermann}, {Aharonson}, {Birnholtz}, {Chelouche}, {Guetta}, {Hallakoun},
  {Horesh}, {Kushnir}, {Mazeh}, {Nordin}, {Ofir}, {Ohm}, {Parsons}, {Pe'er},
  {Perets}, {Perdelwitz}, {Poznanski}, {Sadeh}, {Sagiv}, {Shahaf}, {Soumagnac},
  {Tal-Or}, {Van Santen}, {Zackay}, {Guttman}, {Rekhi}, {Townsend},
  {Weinstein}, \& {Wold}}]{ultrasat}
{Shvartzvald}, Y., {Waxman}, E., {Gal-Yam}, A., {et~al.} 2023{\natexlab{a}},
  arXiv e-prints, arXiv:2304.14482, \dodoi{10.48550/arXiv.2304.14482}

\bibitem[{{Shvartzvald} {et~al.}(2023{\natexlab{b}}){Shvartzvald}, {Waxman},
  {Gal-Yam}, {Ofek}, {Ben-Ami}, {Berge}, {Kowalski}, {B{\"u}hler}, {Worm},
  {Rhoads}, {Arcavi}, {Maoz}, {Polishook}, {Stone}, {Trakhtenbrot},
  {Ackermann}, {Aharonson}, {Birnholtz}, {Chelouche}, {Guetta}, {Hallakoun},
  {Horesh}, {Kushnir}, {Mazeh}, {Nordin}, {Ofir}, {Ohm}, {Parsons}, {Pe'er},
  {Perets}, {Perdelwitz}, {Poznanski}, {Sadeh}, {Sagiv}, {Shahaf}, {Soumagnac},
  {Tal-Or}, {Van Santen}, {Zackay}, {Guttman}, {Rekhi}, {Townsend},
  {Weinstein}, \& {Wold}}]{ultrasat23}
---. 2023{\natexlab{b}}, arXiv e-prints, arXiv:2304.14482,
  \dodoi{10.48550/arXiv.2304.14482}

\bibitem[{Spitkovsky(2006)}]{spitkovsky2006time}
Spitkovsky, A. 2006, The Astrophysical Journal, 648, L51

\bibitem[{Stella {et~al.}(2005)Stella, Dall’Osso, Israel, \&
  Vecchio}]{stella2005gravitational}
Stella, L., Dall’Osso, S., Israel, G., \& Vecchio, A. 2005, The Astrophysical
  Journal, 634, L165

\bibitem[{Sur \& Haskell(2021)}]{sur2021gravitational}
Sur, A., \& Haskell, B. 2021, Monthly Notices of the Royal Astronomical
  Society, 502, 4680

\bibitem[{Thompson \& Duncan(1993)}]{thompson1993neutron}
Thompson, C., \& Duncan, R.~C. 1993, The Astrophysical Journal, 408, 194

\bibitem[{{Woosley}(2010)}]{Woosely2010}
{Woosley}, S.~E. 2010, \apjl, 719, L204, \dodoi{10.1088/2041-8205/719/2/L204}

\end{thebibliography}

\end{document}